\documentclass[aps,prc,twocolumn,superscriptaddress,showpacs,showkeys]{revtex4}
\usepackage{hyperref}

\usepackage{graphicx}
\usepackage{dcolumn}
\usepackage{bm}

\usepackage{float}

\usepackage[usenames]{color}

\begin{document}


\title{Variation of nonequilibrium processes in p+Ni system with
beam energy}



\author{A.Budzanowski}
\affiliation{H. Niewodnicza{\'n}ski Institute of Nuclear Physics
PAN, Radzikowskiego 152, 31342 Krak{\'o}w, Poland}
\author{M.Fidelus}
\affiliation{M. Smoluchowski Institute of Physics, Jagellonian
University, Reymonta 4, 30059 Krak{\'o}w, Poland}
\author{D.Filges}
\affiliation{Institut f{\"u}r Kernphysik, Forschungszentrum
J{\"u}lich, D-52425 J{\"u}lich, Germany}
\author{F.Goldenbaum}
\affiliation{Institut f{\"u}r Kernphysik, Forschungszentrum
J{\"u}lich, D-52425 J{\"u}lich, Germany}
\author{H.Hodde}
\affiliation{Institut f{\"ur} Strahlen- und Kernphysik, Bonn
University,  D-53121 Bonn, Germany}
\author{L.Jarczyk}
\affiliation{M. Smoluchowski Institute of Physics, Jagellonian
University, Reymonta 4, 30059 Krak{\'o}w, Poland}
\author{B.Kamys}  \email[Electronic address: ]{ufkamys@cyf-kr.edu.pl}
\affiliation{M. Smoluchowski Institute of Physics, Jagellonian
University, Reymonta 4, 30059 Krak{\'o}w, Poland}
\author{M.Kistryn}
\affiliation{H. Niewodnicza{\'n}ski Institute of Nuclear Physics
PAN, Radzikowskiego 152, 31342 Krak{\'o}w, Poland}
\author{St.Kistryn}
\affiliation{M. Smoluchowski Institute of Physics, Jagellonian
University, Reymonta 4, 30059 Krak{\'o}w, Poland}
\author{St.Kliczewski}
\affiliation{H. Niewodnicza{\'n}ski Institute of Nuclear Physics
PAN, Radzikowskiego 152, 31342 Krak{\'o}w, Poland}
\author{A.Kowalczyk}
\affiliation{M. Smoluchowski Institute of Physics, Jagellonian
University, Reymonta 4, 30059 Krak{\'o}w, Poland}
\author{E.Kozik}
\affiliation{H. Niewodnicza{\'n}ski Institute of Nuclear Physics
PAN, Radzikowskiego 152, 31342 Krak{\'o}w, Poland}
\author{P.Kulessa}
\affiliation{H. Niewodnicza{\'n}ski Institute of Nuclear Physics
PAN, Radzikowskiego 152, 31342 Krak{\'o}w, Poland}
\affiliation{Institut f{\"u}r Kernphysik, Forschungszentrum
J{\"u}lich, D-52425 J{\"u}lich, Germany}
\author{H.Machner}
\affiliation{Institut f{\"u}r Kernphysik, Forschungszentrum
J{\"u}lich, D-52425 J{\"u}lich, Germany}
\author{A.Magiera}
\affiliation{M. Smoluchowski Institute of Physics, Jagellonian
University, Reymonta 4, 30059 Krak{\'o}w, Poland}
\author{B.Piskor-Ignatowicz}
\affiliation{M. Smoluchowski Institute of Physics, Jagellonian
University, Reymonta 4, 30059 Krak{\'o}w, Poland}
\affiliation{Institut f{\"u}r Kernphysik, Forschungszentrum
J{\"u}lich, D-52425 J{\"u}lich, Germany}
\author{K.Pysz}
\affiliation{H. Niewodnicza{\'n}ski Institute of Nuclear Physics
PAN, Radzikowskiego 152, 31342 Krak{\'o}w, Poland}
\affiliation{Institut f{\"u}r Kernphysik, Forschungszentrum
J{\"u}lich, D-52425 J{\"u}lich, Germany}
\author{Z.Rudy}
\affiliation{M. Smoluchowski Institute of Physics, Jagellonian
University, Reymonta 4, 30059 Krak{\'o}w, Poland}
\author{R.Siudak}
\affiliation{H. Niewodnicza{\'n}ski Institute of Nuclear Physics
PAN, Radzikowskiego 152, 31342 Krak{\'o}w, Poland}
\affiliation{Institut f{\"u}r Kernphysik, Forschungszentrum
J{\"u}lich, D-52425 J{\"u}lich, Germany}
\author{M.Wojciechowski}
\affiliation{M. Smoluchowski Institute of Physics, Jagellonian
University, Reymonta 4, 30059 Krak{\'o}w, Poland}

\collaboration{PISA - \textbf{P}roton \textbf{I}nduced
\textbf{S}p\textbf{A}llation collaboration}

\date{\today}

\begin{abstract}
The energy and angular dependence of double differential cross
sections $d^{2}\sigma/d\Omega dE$ were measured for
$p,d,t,^{3,4}$He, $^{6,7}$Li, $^{7,9}$Be, and $^{10,11}$B produced
in collisions of  0.175 GeV protons with Ni target. The analysis of
measured differential cross sections allowed to extract total
production cross sections for ejectiles listed above. The shape of
the spectra and angular distributions indicate the presence of other
nonequilibrium processes besides the emission of nucleons from the
intranuclear cascade, and besides the evaporation of various
particles from remnants of intranuclear cascade. These
nonequilibrium processes consist of coalescence of nucleons into
light charged particles during the intranuclear cascade, of the
fireball emission which contributes to the cross sections of protons
and deuterons, and of the break-up of the target nucleus which leads
to the emission of intermediate mass fragments. All such processes
were found earlier at beam energies 1.2, 1.9, and 2.5 GeV for Ni as
well as for Au targets, however, significant differences in
properties of these processes at high and low beam energy are
observed in the present study.

\end{abstract}

\pacs{25.40.-h,25.40.Sc,25.40.Ve}

\keywords{Proton induced reactions, production of light charged
particles and intermediate mass fragments, spallation,
fragmentation, nonequilibrium processes, coalescence, fireball
emission}

\maketitle


\section{\label{sec:introduction} Introduction}


One of the most  significant  questions  to be  addressed  by
studies on proton-nucleus collisions is the predictive power of
existing models and computer programs used for their realization.
The above question is closely related to two following problems: (i)
whether all important physical phenomena are taken into
consideration, (ii) whether the parameters of the models are
adjusted properly.  It is well known that neglecting of an important
physical phenomenon may be usually "repaired" in a specific case by
appropriate adjusting of free parameters of the model.  This
procedure cannot be, however, extended to a full set of observables
for all targets and energies. Thus, a general model must explicitly
contain all important physical phenomena.

The traditionally used description of proton induced reactions at
GeV energies assumes  that the reactions proceed in two steps.  The
fast, nonequilibrium step in such two-step model consists in an
intranuclear cascade of nucleon-nucleon collisions with a possible
coalescence of the nucleons into complex particles as it is
realized,e.g., by the INCL4.3 computer program of Boudard \emph{et
al.} \cite{BOU04A}. This stage of the reaction is assumed to lead to
an equilibrated, excited residuum of the target, which in the
following evaporates particles or/and emits fission fragments. This
picture of reactions turned out to be realistic in many situations.
It was, however, observed at proton beam energies above several GeV
that copious emission of intermediate mass fragments (IMFs), i.e.,
ejectiles heavier than $^4$He and lighter than fission fragments,
appears (cf., e.g., \cite{RIC01A,VIO06A}) what is interpreted as an
analogue of the liquid - gas phase transition (cf., e.g.,
\cite{KAR03A,KAR04A} and references therein). The nucleus, which is
treated as a liquid, changes then into a mixture of free nucleons,
light charged particles -- LCPs (particles with Z $\leq$ 2) and
IMFs, treated as a fog. In this case, ejectiles are emitted by only
one source —- the slowly moving target spectator.

It was, however, recently found that at proton beam energies 1.2 -
2.5 GeV the IMFs as well as LCPs originating from p+Ni \cite{BUD09A}
and p+Au  collisions \cite{BUB07A,BUD08A}  are emitted from three
sources. They are interpreted as a fireball - fast and hot source
consisted of several nucleons - knocked out by the impinging proton,
and two slower and colder sources which are believed to be
prefragments of the target nucleus appearing due to its break-up
caused by strong deformation induced by the fireball emission. The
analysis of the experimental data by a traditional model assuming
the presence of intranuclear cascade with the possibility to form
complex particles due to coalescence, and evaporation from an
equilibrated target remnant could not take account for the presence
of these sources and could not reproduce  the full set of
experimental data. On the contrary, the combination of a traditional
model with additional inclusion of the emission from the fireball
and two other sources, treated within a phenomenological model, led
to a perfect description of energy and angular dependencies of all
double differential cross sections $d^2\sigma/d\Omega dE$. It is
therefore obvious, that in a proper theoretical analysis such
phenomena should be taken into consideration.

The question arises on the energy development of the reaction
mechanism.  It is not clear, whether the same picture of the
reaction may be applied to other energies - below and above the
studied proton beam energy range; 1.2 - 2.5 GeV.  In the present
study the investigation of the reaction mechanism of p+Ni collisions
is extended to much lower energy E$_p$=0.175 GeV, which is on the
boarder of applicability of the traditional model of an intranuclear
cascade followed by an evaporation \cite{CUG87A,CUG97A,BOU04A}.

To facilitate the comparison of the results from the present study
of the reactions in p+Ni system with results of previous
investigations at higher energies in the same nuclear system
\cite{BUD09A}, the present paper is organized in a similar way as
Ref. \cite{BUD09A}. Experimental data are discussed in the next
section, the theoretical analysis is described in the third section
starting from IMF data and followed by the analysis of LCPs cross
sections, the discussion of results is presented in the fourth and
the summary with conclusions in the fifth section.


\section{\label{sec:experiment}Experimental data}

The experiment was performed with a selfsupporting Ni target of the
thickness of about 150 $\mu$g/cm$^{2}$, irradiated by an internal
proton beam of COSY (COoler SYnchrotron) of the J\"ulich Research
Center. The experimental setup and procedure of data taking were in
details described in Refs. \cite{BUB07A} and \cite{BAR04A}.

Double differential cross sections $d^{2}\sigma/d\Omega dE$ were
measured at seven scattering angles; 16$^0$, 20$^0$, 35$^0$, 50$^0$,
65$^0$, 80$^0$, and 100$^0$ as a function of energy of ejectiles for
the following isotopes $^{1,2,3}$H, $^{3,4}$He, $^{6,7}$Li,
$^{7,9}$Be, and $^{10,11}$B.

\begin{figure}
\begin{center}
\includegraphics[width=8cm,angle=0]{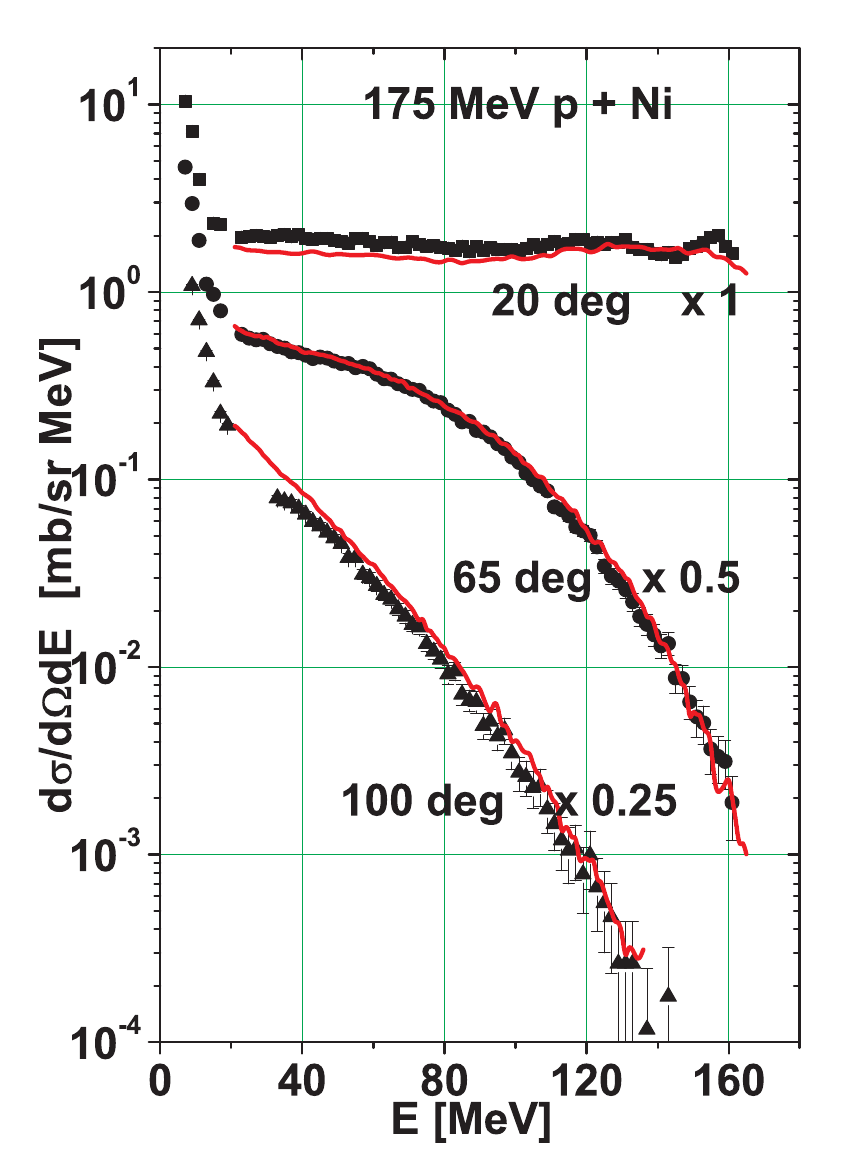}
\caption{Proton energy spectra measured at 20$^0$, 65$^0$, and
100$^0$ in the laboratory system. Lines represent the data from
F\"ortsch \emph{et al.} \cite{FOR91A}, symbols depict the data from
the present experiment. The spectra were multiplied by factors
written in the figure to avoid overlapping the symbols and lines
obtained at different angles. } \label{fig:normalization}
\end{center}
\end{figure}

The absolute normalization of the cross sections was achieved by
comparing the proton differential cross sections measured in the
present experiment at 20$^0$, 65$^0$, and 100$^0$ with the
absolutely normalized proton spectra from the experiment of
F\"ortsch \emph{et al.} \cite{FOR91A}. A perfect agreement of the
shape of the spectra from both experiments as well as an agreement
of their angular dependence can be seen in Fig.
\ref{fig:normalization}.  It is worthy to point out, that the
spectra consist of two parts:  a low energy part (energy smaller
than $\sim$ 20 MeV) - measured only in the present experiment, where
the evaporation of protons from excited target remnants after the
intranuclear cascade sets in, and high energy tails, which are due
to preequilibrium processes. In the traditional, two-step model this
part of the spectra is due to the emission of protons  from the
intranuclear cascade. As it will be discussed below, the same two
components - representing particles emitted from the equilibrated
nuclear system as well as those from preequilibrium processes - are
visible in the spectra of other LCPs and IMFs.

\section{\label{sec:analysis} Theoretical analysis}

The analysis of present experimental data was performed according to
the same procedure as that applied previously to the data from
proton induced reactions on Ni target in the work of Budzanowski
\emph{et al.} \cite{BUD09A} at higher energies.

The experimental data were first compared with calculations
performed in the frame of two step model in which the fast stage was
calculated as intranuclear cascade with the possibility to coalesce
the outgoing nucleons into complex LCPs, and the slow stage was
modeled by evaporation of particles (both LCPs and IMFs) from the
excited residuum of the intranuclear cascade, which was assumed to
be in equilibrium.

\begin{figure}
\begin{center}
\hspace*{-0.5cm}
\includegraphics[angle=0,width=0.53\textwidth]{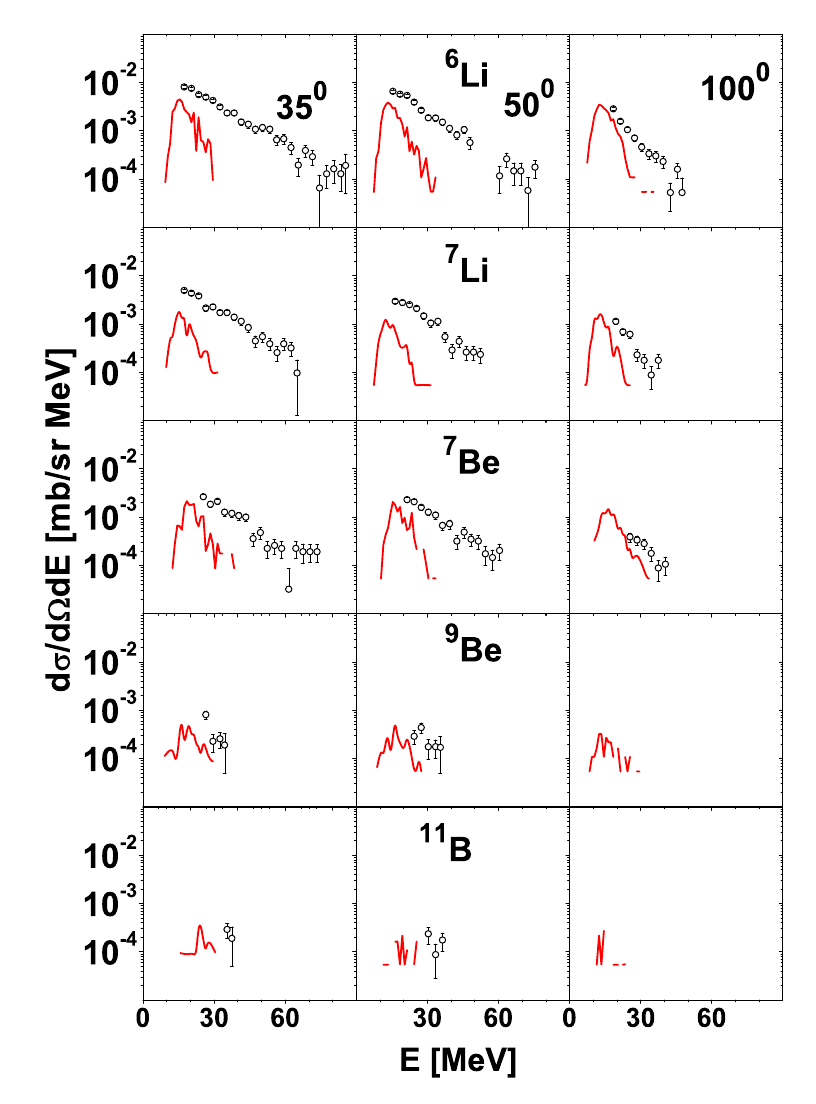}
\caption{\label{fig:libebinc50} Typical spectra of selected lithium,
beryllium, and boron isotopes from p+Ni collisions measured at
35$^{\circ}$, 50$^{\circ}$, and 100$^{\circ}$ (left, middle, and
right columns of the figure, respectively)  for 0.175 GeV proton
beam impinging on to the Ni target. The detected particles are
listed in the central panel of each row. Open circles represent the
experimental data,  and solid lines correspond to intranuclear
cascade followed by evaporation of particles, respectively.}
\end{center}
\end{figure}

\begin{figure}
\begin{center}
\hspace*{-0.5cm}
\includegraphics[angle=0,width=0.54\textwidth]{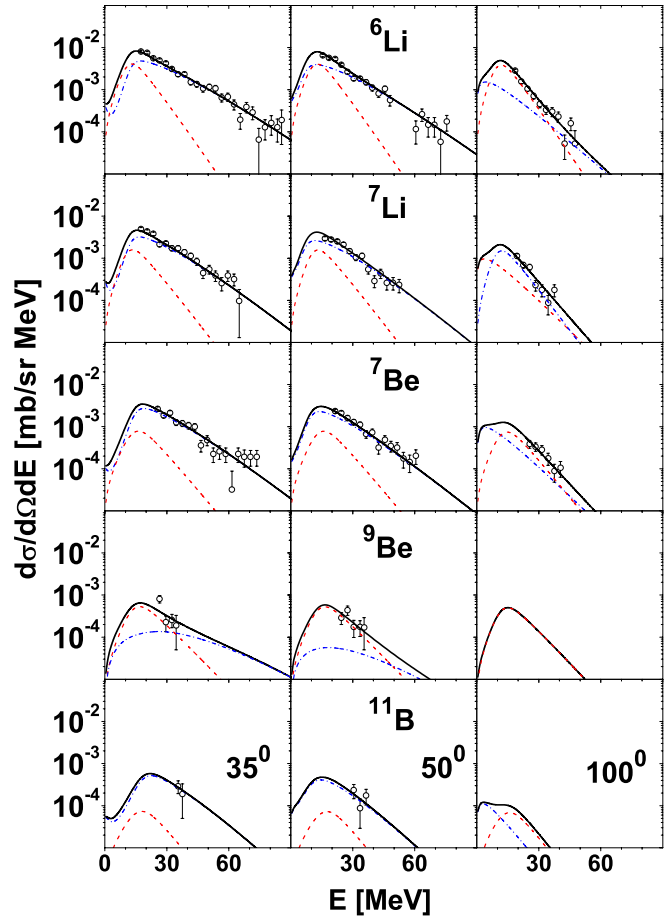}
\caption{\label{fig:libeb} 
Same as Fig. \ref{fig:libebinc50} but dashed, dot - dashed, and
solid lines correspond to slow emitting source, fast emitting source
and the sum of both contributions, respectively. }
\end{center}
\end{figure}

The calculations of the first step of the reaction were done using
the INCL4.3 computer program of Boudard \emph{et al.} \cite{BOU04A},
and the calculations of evaporation of particles were realized by
means of GEM2 computer program of Furihata \cite{FUR00A,FUR02A}.  In
both types of calculations default values of the parameters,
proposed by authors were used, thus no adjusting of the theoretical
cross sections to the data was undertaken.

It turned out, that the spectra of both, LCPs and IMFs were not
satisfactorily well reproduced.  Therefore a phenomenological
analysis was performed in which the isotropic emission of particles
from sources moving in forward direction (along to the beam) was
allowed. Each of the sources had Maxwellian distribution of the
energy $E$ available for the two body decay in which the emission of
the detected particles occured; $d^2 \sigma/dE d\Omega \sim \sqrt{E}
\exp(-E/T)$. The velocity of the source - $\beta$ (in units of speed
of light), its temperature - $T$ (in MeV), and the contribution to
the total production cross section - $\sigma$ (in mb) were treated
as free parameters.

%
\begin{table*}
  \caption{\label{tab:parameters}Parameters of two moving sources for isotopically identified
  IMF's and for $^4$He: $\beta_i$, $T_i$, and $\sigma_i$ correspond to  source velocity, its apparent temperature,
 and the total production cross section, respectively. The sum $\sigma \equiv \sigma_1+\sigma_2$ is
 also listed.  The left part of the Table
 (parameters with indices "1")
 corresponds to the slow moving source, and the right part  contains values of parameters for the fast moving source.
}
\begin{tabular}{lrr|llr|ccr}
  \hline \hline
            & \multicolumn{2}{c|}{Slow source} & \multicolumn{3}{c|}{Fast source}                 &             & \multicolumn{2}{r}{} \\
  \cline{2-6}
  Ejectile   & \emph{T}$_1$/MeV & $\sigma_1$/mb & \hspace*{0.5cm} $\beta_2$   & \emph{T}$_2$/MeV & $\sigma_2$/mb & \hspace*{0.5cm} $\sigma$/mb  & $\sigma_2/\sigma$   & $\chi^2$ \\
  \hline
  \hline
  $^4$He     &  7.0(3)          &  39(3)        & \hspace*{0.5cm} 0.060(6)    & 9.7(4 )          &  16.8(2.7)    & \hspace*{0.5cm} 55.8(4.2)    & 0.30(6)            & 32.5\\
  $^6$Li     &  5.7(4)          &  0.70(6)      & \hspace*{0.5cm} 0.045(2)    & 10.0(2)          &  0.71(5)      & \hspace*{0.5cm} 1.41(8)      & 0.50(5)  & 1.4\\
  $^7$Li     &  6.3(8)          &  0.29(5)      & \hspace*{0.5cm} 0.041(2)    &  8.7(4)          &  0.41(4)      & \hspace*{0.5cm} 0.70(7)      & 0.59(8)  & 1.4\\
  $^7$Be     &  6.8(1.8)        &  0.16(7)      & \hspace*{0.5cm} 0.040(4)    &  9.0(6)          &  0.37(6)      & \hspace*{0.5cm} 0.53(9)      & 0.70(16) & 1.2\\
  $^9$Be     &  [6.5]           &  0.12(6)      & \hspace*{0.5cm} 0.08(2)     &  [9.0]           &  0.02(1)      & \hspace*{0.5cm} 0.14(6)      & 0.14(7)  & 1.4\\
  $^{10}$B   &  [6.5]            & 0.10(9)      & \hspace*{0.5cm} [0.04]      &  7.1(3.8)        &  0.07(4)      & \hspace*{0.5cm} 0.17(10)    & 0.42(34) & 1.7\\
  $^{11}$B   &  [6.5]            & 0.020(14)    & \hspace*{0.5cm} [0.04]      &  7.0(7.3)        &  0.06(4)      & \hspace*{0.5cm} 0.08(5)     & 0.75(69) &  1.3\\
  \hline
  \hline
\end{tabular}
\end{table*}
%

Two additional parameters, defining the height $B$ of the Coulomb
barrier for emitted particles and the diffuseness $d$ of the
transmission function through the barrier, were used with fixed
values. Further details of the model are described in the Appendix
of  Ref. \cite{BUB07A}.

The analysis of IMF data differs from that of LCP cross sections,
thus they are described separately below.

\subsection{\label{sec:IMF} Intermediate mass fragments}

The experimental spectra for $^{6,7}$Li, $^{7,9}$Be, and $^{11}$B
measured at 35$^{\circ}$, 50$^{\circ}$, and 100$^{\circ}$ scattering
angles are shown in Fig. \ref{fig:libebinc50} as open dots together
with the theoretical calculations performed in the frame of the
two-step model (intranuclear cascade followed by evaporation of
particles from the excited remnant of the target nucleus), which are
depicted by solid lines.  The fluctuations of these lines are due to
limited statistics of model calculations which were done by the
Monte Carlo method and have no physical meaning.

As can be seen the theoretical spectra are different from the
experimental ones in several subjects; (i) the theoretical spectra
are almost independent of the scattering angle, whereas the
experimental spectra vary with the angle showing increasing of the
slope with the scattering angle, (ii) the theoretical spectra are
localized at small ejectile energies ( smaller than $\sim$ 30 MeV )
whereas the experimental spectra cover much larger range of
energies, especially for small scattering angles, (iii) the
magnitude of the theoretical cross sections is smaller than that of
the data.

In the second step of the analysis the emission of IMFs from two
moving sources has been calculated adding the cross sections from
both sources.  The velocity of the fast source $\beta_2$,
temperature parameters of both sources $T_1$ and $T_2$, and total
production cross sections $\sigma_1$ and $\sigma_2$ due to both
sources were fitted to obtain the best agreement of the theoretical
cross sections with the data for all seven scattering angles
simultaneously. The other parameters, i.e., velocity of the slow
source $\beta_1$ as well as parameters characterizing the Coulomb
barrier $k_1$, $k_2$ (heights of Coulomb barrier between ejectile
and source in units $B$, i.e. height of Coulomb barrier between
ejectile and the target nucleus) and $(B/d)_1$, $(B/d)_2$ were
fixed. The velocity of the slow source was assumed to be equal to
the average velocity of the residual nuclei after the intranuclear
cascade $\beta_1 = 0.0036$ as it was extracted from INCL4.3
calculations, and Coulomb barrier parameters were fixed at
arbitrarily chosen values $k_1$= 0.75, $k_2=0.3$, and $(B/d)_1 =
(B/d)_2$=5.5. These parameters almost do not influence the spectra
with exception of very low ejectile energies, thus the same values
of the parameters were taken as those used at higher beam energies
\cite{BUD09A}. The experimental spectra measured at 35$^{\circ}$,
50$^{\circ}$, and 100$^{\circ}$ (open circles) are shown in Fig.
\ref{fig:libeb} together with results of the calculations (lines).
The solid line represents the sum of contributions from both
sources, dashed line depicts the cross section originating from the
slow source, and dot-dashed line shows cross section corresponding
to emission from the fast source.

A very good description of the data has been obtained with the
parameters varying smoothly from ejectile to ejectile.  The values
of the parameters are listed in  Table \ref{tab:parameters}. The
errors of the parameters, estimated by a computer program, which
searched for best fit parameters are also given in the Table.  Some
parameters -- closed in square brackets -- were fixed during the fit
to avoid ambiguities of the parameters, which appear when the data
do not put strong enough constraints to the parameters.

As can be seen in Fig. \ref{fig:libeb}, the slow source produces
spectra which are almost independent of angle and are similar to
those calculated from two-step microscopic model.  The fast source
contribution to the spectra is angle dependent thus it represents
the nonequilibrium process proceeding in the fast stage of the
reaction.  The spectrum evaluated for this source has a high-energy
tail which allows to reproduce the high energy part of the
experimental spectra. The relative contribution of this source to
the total production cross section is large as can be checked in the
Table \ref{tab:parameters} (in average it is equal to 53(12)\%).

\begin{figure}
\begin{center}
\vspace*{-0.5cm}
\hspace*{-0.5cm}
\includegraphics[angle=0,width=0.5\textwidth]{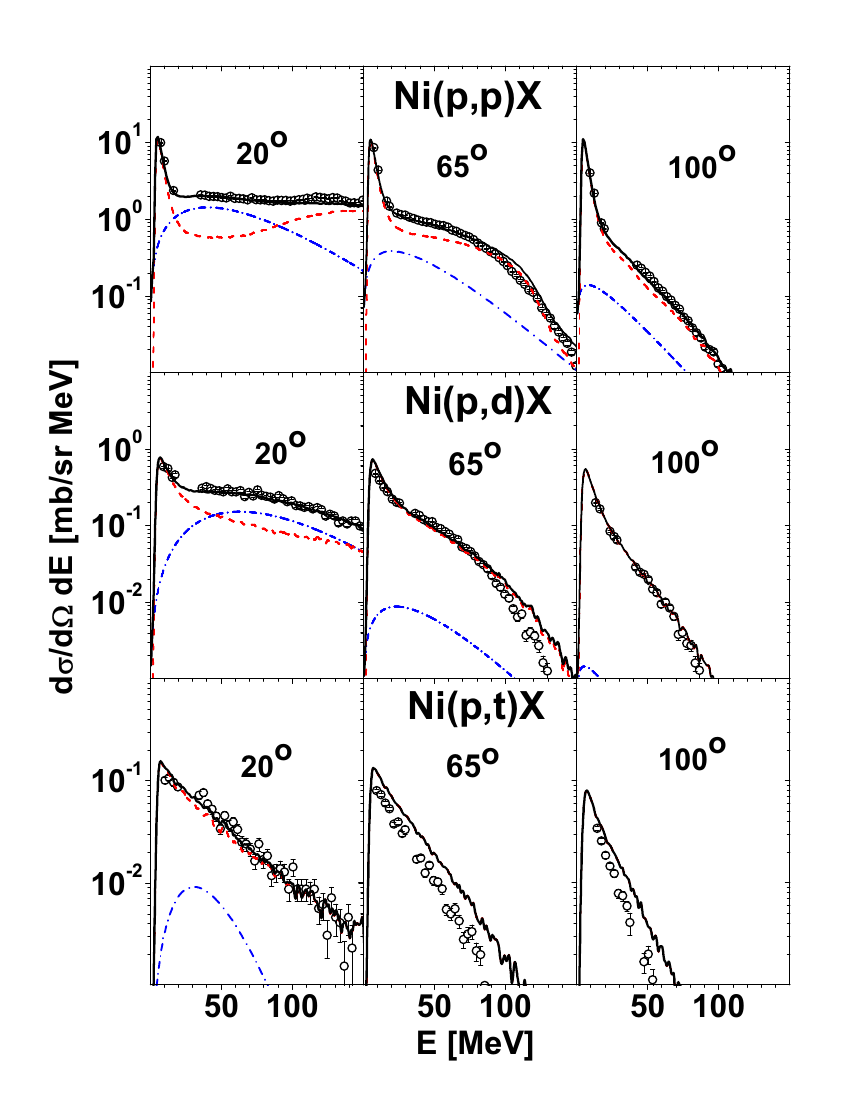}
\caption{\label{fig:pdt} Typical spectra of protons, deuterons, and
tritons (upper, middle, and lower rows of the figure, respectively)
measured at 20$^{\circ}$, 65$^{\circ}$, and 100$^{\circ}$ (left,
middle, and right columns of the figure, respectively)  for 0.175
GeV  proton beam impinging on to the Ni target. Open circles
represent the experimental data, dashed, dot - dashed, and solid
lines correspond to two-step model (scaled by factor F - for
explanation see text), emission from the fireball and sum of both
contributions, respectively. Contribution of the fireball is very
small for deuterons emitted at large angles as well as for tritons
at all scattering angles. }
\end{center}
\end{figure}

\begin{figure}
\begin{center}
\vspace*{-0.5cm}
\hspace*{-0.5cm}
\includegraphics[angle=0,width=0.5\textwidth]{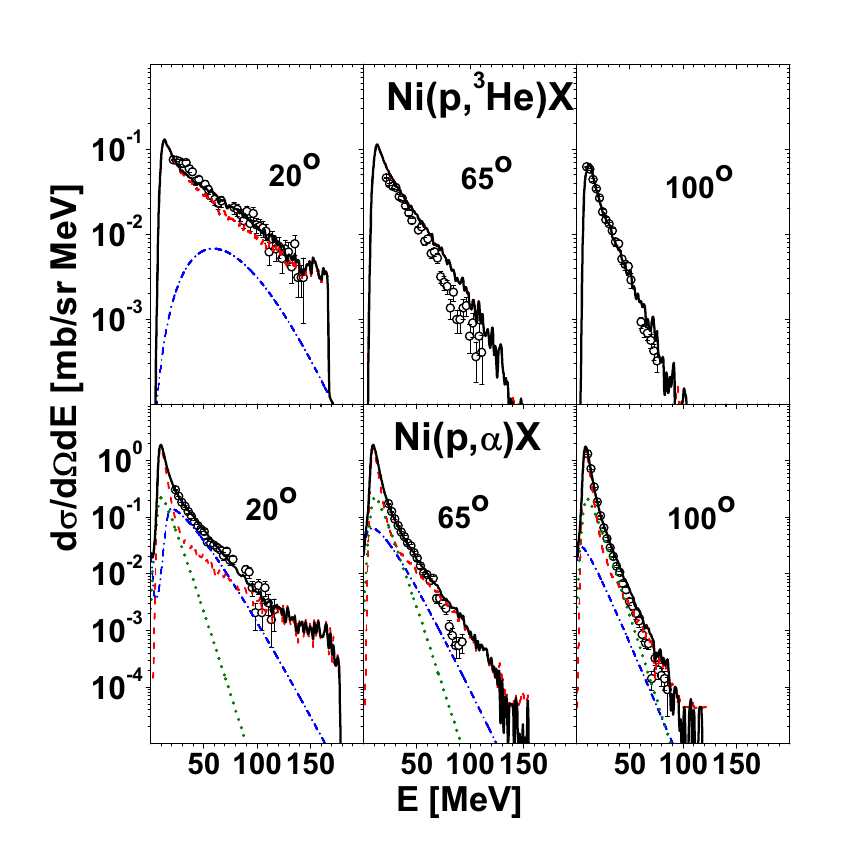}
\caption{\label{fig:he34} Typical spectra of $^3$He and $^4$He
(upper, and lower rows of the figure, respectively) measured at
20$^{\circ}$, 65$^{\circ}$, and 100$^{\circ}$ (left, middle, and
right columns of the figure, respectively)  for 0.175 GeV proton
beam impinging on to the Ni target. Open circles represent the
experimental data, dashed lines represent results of two-step model
(scaled by factor F), and solid line depicts sum of all
contributions. Dot-dashed line in the upper panel shows contribution
of the fireball whereas the dot-dashed and dotted lines for $^4$He
denote contributions of fast and slow moving sources, respectively.
}
\end{center}
\end{figure}

\subsection{\label{sec:LCP} Light charged particles}

The experimental spectra of LCPs extend to energies higher than 20 -
30 MeV, which is the upper limit of energy for evaporated particles.
Thus, it is obvious that the nonequilibrium emission of particles is
responsible for the higher energy part of the spectra.  In the case
of protons, such nonequilibrium emission appears from the
intranuclear cascade before achieving  an equilibrium in the target
residuum.  Coalescence of nucleons of the target with the nucleons
escaping from the intranuclear cascade, which may proceed if the
relative spatial and momentum position of nucleons is small enough,
has been considered as the process responsible for emission of
complex LCPs. Letourneau \emph{et al.} \cite{LET02A} and Boudard
\emph{et al.} \cite{BOU04A} proposed to treat the coalescence
microscopically during the calculation  of an intranuclear cascade.
Thus this phenomenon is implemented in the INCL4.3 computer program
\cite{BOU04A} and therefore  this program has been applied in the
present work for evaluation of intranuclear cascade and coalescence
of nucleons leading to the formation of deuterons, tritons, $^{3}$He
and alpha particles. The results of these calculations, coupled with
the evaporation of particles evaluated by means of the GEM2 computer
program were compared with the experimental spectra. Very good
reproduction of triton and $^3$He spectra was achieved for all
scattering angles as well as significant improvement (in comparison
to evaporation spectra alone) of deuteron and alpha particle spectra
for large scattering angles. However, the small scattering angles of
protons, deuterons and alpha particles were still not satisfactorily
well reproduced. Moreover, it was found that the improvement of the
description of LCPs spectra by inclusion of coalescence deteriorates
simultaneously the description of the proton spectra, because
increasing of the production of composite particles occurs on the
account of decreasing the emission of nucleons.

It was thus assumed that an additional process, namely the emission
of a fireball and break-up of the target nucleus, should be taken
into consideration as it was found to be necessary for p+Ni and p+Au
collisions at higher proton energies (1.2, 1.9, and 2.5 GeV),
investigated by present authors (Refs. \cite{BUD09A} and
\cite{BUD08A}, respectively). The parameters of the fireball, i.e.,
its temperature parameter - $T_3$, velocity of the source -
$\beta_3$, total production cross section associated with this
mechanism - $\sigma_3$ were treated as free parameters and modified
to obtain the best description of experimental cross sections. Other
parameters, i.e., $k_3$ (the height of the Coulomb barriers in units
of $B$ - Coulomb barrier between the ejectile and the target
nucleus) and the parameter $B/d$ describing diffuseness of the
transmission function through the Coulomb barrier were fixed at
arbitrarily assumed values 0.07 and 4.8, respectively. It should be
emphasized, that the coalescence and evaporation cross sections were
allowed to be scaled down by an adjustable factor F, what physically
is understood as making room for new nonequilibrium process, which
in original INCL4.3+GEM2 calculations was not considered. Values of
the fitted parameters are collected in the Table \ref{tab:fireball}.
%
%
\begin{table*} 
 \caption{\label{tab:fireball} Parameters  $\beta_3$, $T_3$,
 and $\sigma_3$ correspond to the fireball velocity in units of speed of light, its apparent temperature,
 and the total production cross section, respectively.
 Parameter F is the scaling factor of coalescence and evaporation
 contribution extracted from fit to the proton spectra and deuteron spectra. The numbers in parentheses
 show fixed values of the parameters.
 The columns described as F$\ast \sigma_{INCL}$ and F$\ast \sigma_{GEM}$ contain total
 production cross sections due to intranuclear cascade with the coalescence and due to evaporation
 from the target residuum, respectively. The total production cross section obtained by summing of all contributions
 is depicted in the column denoted by $\sigma$. In the case of alpha particles it contains also
 the contribution of emission from slow and fast sources listed in  Table \ref{tab:parameters}.}
\begin{center}
 \begin{tabular}{lcrrccrrrr}
\hline \hline
  Ejectile & $\beta_3$  &  $T_3$    &  $\sigma_3$ &     F   & F$\ast \sigma_{INCL}$ & F$\ast \sigma_{GEM}$ & $\sigma$  & $\sigma_3/\sigma$ & $\chi^{2} $ \\
           &            &  MeV      &  mb         &         &    mb             &   mb             &  mb       &                   & \\
\hline
  \emph{p}  & 0.232(5)  & 21.2(1.3) &  320(32)    & \hspace*{0.5cm} 0.83(5) &   567             &    697           & \hspace*{0.5cm} 1584(32)  &   0.20(2)          & 26.8 \\
  \emph{d}  & 0.240(9)  & 16.8(2.0) &  22.9(3.7)  & \hspace*{0.5cm} 0.80(3) &   104             &    31            & \hspace*{0.5cm} 158(5)    &   0.14(3)          & 5.3   \\
  \emph{t}  & 0.142(27) &  6.1(4.9) &    [0.5]    & \hspace*{0.5cm} [0.8] &    24.1            &    2.8           & \hspace*{0.5cm} 27.4      &   0.02   & 14.3 \\
 $^{3}$He   & 0.205(20) &  7.3(3.0) &    [0.5]    & \hspace*{0.5cm} [0.8] &    16.0            &    4.8           & \hspace*{0.5cm} 21.3      &   0.02   & 18.8 \\
 $^{4}$He  &            &           &             & \hspace*{0.5cm} [0.8] &    11.5            &    129           & \hspace*{0.5cm} 196.3(4.2) &         & 32.5 \\
\hline \hline
\end{tabular}
\end{center}
 \end{table*}
%
%


It turned out that the inclusion of the emission of LCPs from the
fireball into the analysis leads to a very good description of
proton and deuteron spectra but it gives only negligible
modification of triton and $^3$He theoretical spectra as it is
visible in Figs. \ref{fig:pdt} and \ref{fig:he34}.  The importance
of the fireball contribution to proton and deuteron data may be also
judged from ratio of total production cross section of these
particles via fireball emission to cross section representing sum of
all processes. As can be seen in Table \ref{tab:fireball} the
relative fireball contribution to proton and to deuteron cross
sections is equal to 20(2)\% and 14(3)\% , respectively. This is a
rather small value in spite of the fact that it is crucial for a
proper description of the spectra, especially at forward angles. On
the contrary, the fireball is practically negligible as concerns the
total production cross section of tritons and $^3$He particles.  Its
contribution is of order of 2\% only.  It should be pointed out that
the  scaling factor F of INCL4.3 and GEM2 cross section was fixed
for tritons, $^3$He, and $4$He at the same value as for deuterons,
i.e., F=0.8. Some smaller value will perhaps improve the description
of the cross sections, since theoretical cross sections of INCL4.3
and GEM2 slightly overestimate the data, however the contribution of
the fireball to triton and $^3$He spectra is very small and thus
searching F and fireball cross section $\sigma_3$ independently led
to ambiguities of parameters.

The alpha particle cross sections need an additional contribution
from two moving sources besides the coalescence and evaporation
cross sections.  The parameters of these sources are different from
fireball parameters but are quite similar to those found earlier for
IMFs. It can be thus stated, that the alpha particles behave more as
IMFs than as LCPs.  The same effect has been observed at beam
energies over 1 GeV for p+Ni system \cite{BUD09A}.

\begin{figure}
\begin{center}
\vspace*{-0.7cm}
\includegraphics[angle=0,width=0.5\textwidth]{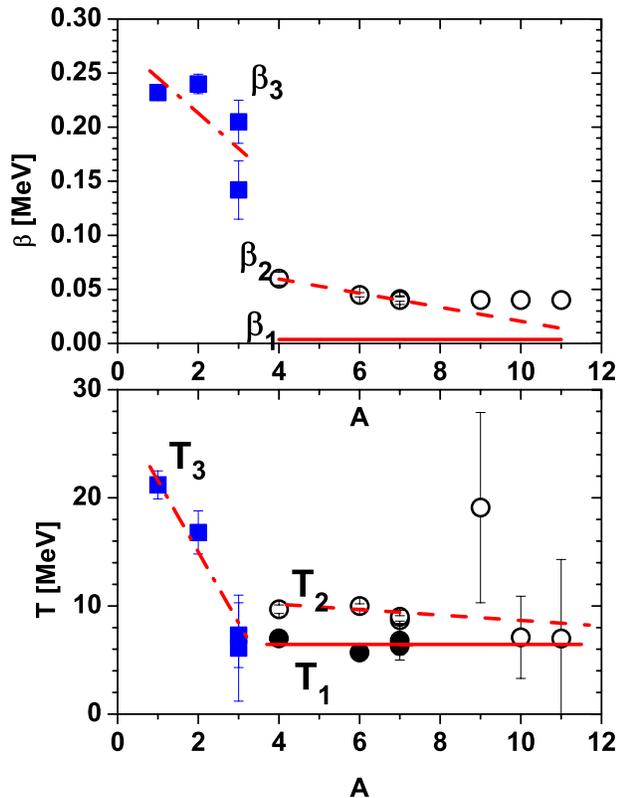}
\caption{\label{fig:bt123L} In the lower panel of the figure the
apparent temperature of the moving sources is drawn as a function of
the ejectile mass A. Open circles and full dots represent values of
temperature parameters $T_2$ and $T_1$ for fast and slow source,
respectively. Full squares indicate temperature $T_3$ of the
fireball.  The solid and dashed lines were fitted to the points
representing the lightest IMF's: $^6$Li, $^7$Li, $^7$Be, and $^4$He.
Dash dotted line was fitted to points representing the LCP's. In the
upper panel of the figure the dependence of the velocity of the
sources is drawn versus the mass of ejectiles. The symbols and lines
have the same meaning as for the lower part of the figure with one
exception: The full dots are not shown because the velocity of the
slow source was fixed during the analysis (at velocity
$\beta_1$=0.0036) and it is represented by the solid line in the
figure. }
\end{center}
\end{figure}

\section{\label{sec:discussion}Discussion}

The velocity and temperature parameters of moving sources for all
ejectiles are presented on Fig. \ref{fig:bt123L} in function of the
mass of ejectiles. Their values behave in very similar manner as it
was observed at higher beam energies in p+Ni \cite{BUD09A} and p+Au
\cite{BUD08A} systems, i.e., they belong to three well separated
sets, representing the slow source ($\beta_1$ and $T_1$),  the fast
source ($\beta_2$ and $T_2$), and the fireball ($\beta_3$ and
$T_3$). The ejectile mass dependence may be approximated by straight
line for each source with a slope which is larger for the fireball
than for the fast source. A velocity of the source can be also
dependent on the mass of the ejectile because mass of the source may
vary for different ejectiles.

   As it was discussed in the previous
paper dealing with reactions in p+Ni system at higher energies
\cite{BUD09A}, the linear dependence of the temperature parameter on
the mass of ejectiles can be used for an estimation of mass of the
source. The parameters of linear functions describing dependence of
the temperature parameter $T$ and velocity $\beta$ of three sources
on the ejectile mass  A are collected in  Table \ref{tab:lowhigh}.
The velocity of the slow source has been fixed at 0.0036 c, i.e., at
an average velocity of residua of intranuclear cascade extracted
from INCL4.3 calculations.  The temperature of this source was also
found to be independent of the mass of ejectile and may be
represented by its average value $\sim$ 6.5 MeV.

The  small and not well defined slope of the mass dependence of the
temperature for the fast source and the fireball give very crude
estimation of their masses, i.e., 45(33) and 4.2(1.0) mass units,
respectively.
The former mass value has too large error to be used for any further
reasoning. The latter value, apart from being not well determined,
is smaller than the mass of the fireball extracted from high energy
data, i.e., 5.5(3) mass units \cite{BUD09A}.  Such decreasing of the
mass of the fireball seems to be in accord with the fact that only
proton and deuteron spectra have significant contribution of
emission from the fireball at 0.175 GeV beam energy, whereas at
higher energies such a contribution was quite large also for tritons
and $^3$He particles.

The next important difference concerns the values of temperature
parameters.  The temperature parameters at low beam energy are
approximately two times smaller than appropriate parameters
determined at high energies. This seems to be a natural consequence
of the fact that the  excitation energy of the target increases with
the beam energy since, at approximately the same rate of the energy
transfer, the total transferred energy must be larger at higher beam
energy.  The same reasoning can be applied to the increase of the
momentum transfer from projectile to the target, which mainly
determines the velocity of the target residuum after the
intranuclear cascade.  According to intranuclear cascade
calculations the velocity of target residuum at low beam energy,
i.e. 0.175 GeV is equal to 0.0036 c, whereas at high energies it is
larger, i.e., equal to 0.0050 c, and is almost independent of energy
in the range 1.2 - 2.5 GeV.

The interesting observation, contradicting to the above reasoning
 is, that \emph{the velocities of the fast source as well as
of the fireball found from the fit are significantly larger at 0.175
GeV beam energy than at higher beam energies}.  This may indicate,
that properties of the reaction mechanism at low energy are
different than those at 1.2 - 2.5 GeV.

According to the present
phenomenological model the sources of ejectiles move along the beam,
thus the momentum conservation requires that the \emph{algebraic}
sum of momenta of all sources should be equal to
projectile momentum ($p_b$=0.60 GeV/c). 
Assuming that the fitted values of the fireball velocity $\beta_3$
for p, d, t, and $^3$He are not biased by errors it is possible to
estimate the maximal mass of the fireball -- 2.7 mass units -- which
assures that the momentum of the fireball  emitting all these
particles is not larger than the beam momentum. Assuming the mass of
the fireball not larger than 3 mass units excludes the possibility
of emission of tritons and $^3$He, what is compatible with the fact
that contribution of the fireball found from the fit to spectra of
these particles  is negligibly small. Furthermore, the emission of
protons and deuterons is possible from such a three-nucleon fireball
but then their momenta evaluated with velocities $\beta_3$ of Table
\ref{tab:fireball} exhaust full available momentum:
$p(fireball_p)$=0.65~GeV/c, $p(fireball_d)$=0.68~GeV/c. The quoted
above momenta are even slightly larger than the beam momentum $p_b$,
but the fact that the fireball velocity was found from unconstraint
fit and is biased by errors, the agreement of momenta of the
fireballs emitting protons and deuterons with the beam momentum is
quite good. Thus, presence of 3-nucleon fireball emitting protons
and/or deuterons is in agreement with momentum conservations and
shows that \emph{full beam momentum is transferred to the fireball}.
This specifically means that: (i) \emph{The proton from the beam
cannot fly away separately from the fireball}, (ii) \emph{the
creation of a fireball cannot simultaneously lead to break-up of the
rest of the target}.

The fact, that two sources, heavier than the fireball are observed
in the analysis of IMFs means that a break-up of the target occurs.
Since the fireball emission cannot be accompanied by the break-up,
\emph{the break-up must proceed without fireball emission}.
%
%
\begin{figure}
\begin{center}
\includegraphics[angle=0,width=0.5\textwidth]{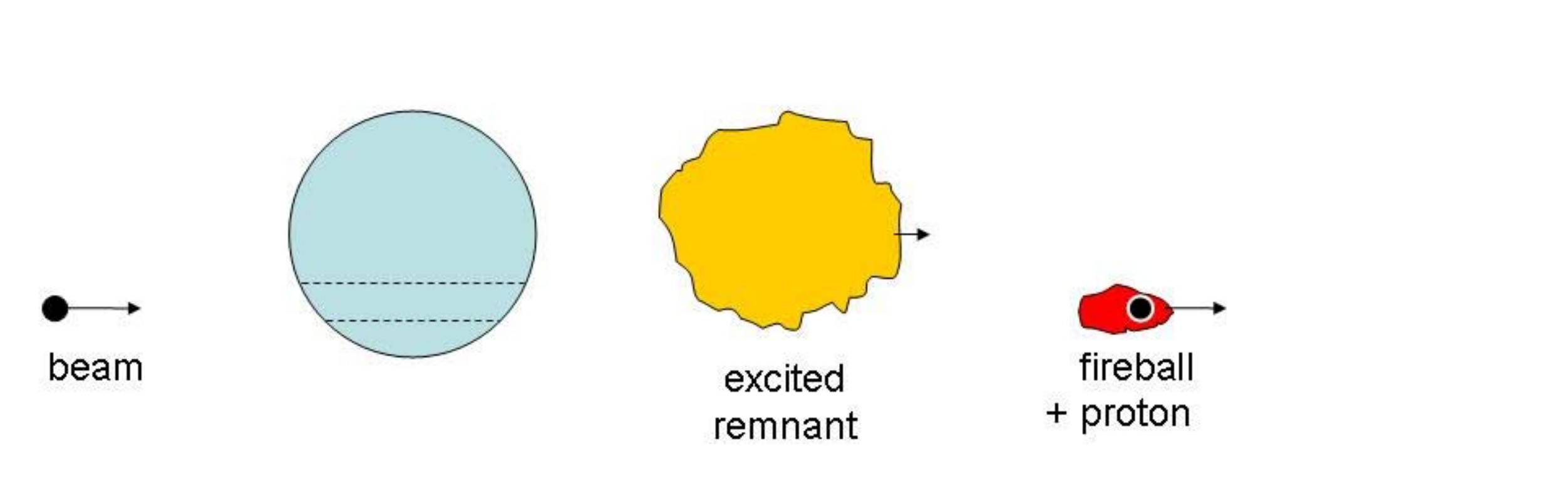}
\caption{\label{fig:firelow} Fireball emission from Ni target at
0.175 GeV beam energy.}
\end{center}
\end{figure}
\begin{figure}
\begin{center}
\includegraphics[angle=0,width=0.5\textwidth]{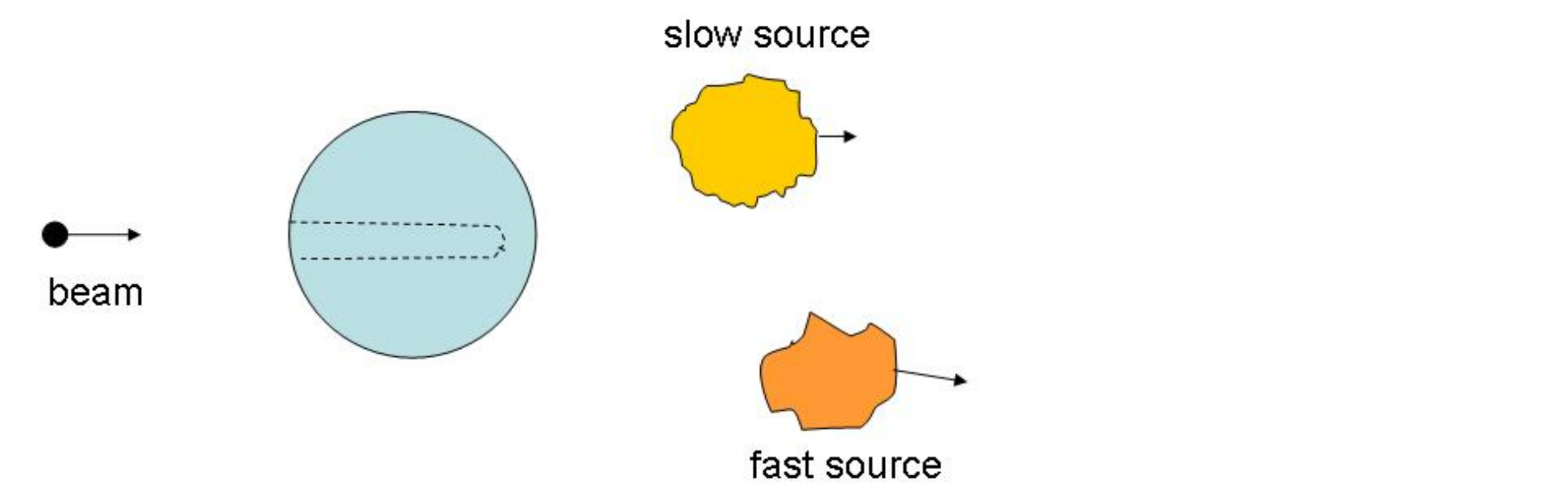}
\caption{\label{fig:break} Break-up of Ni target at 0.175 GeV beam
energy.}
\end{center}
\end{figure}
It is worthy to mention, that such a capture of the projectile
leading to excitation of the nucleus without emission of the
fireball, has been discussed by Aichelin et al. \cite{AIC84A}. They
estimated  that protons of energies smaller than 0.2 - 0.25 GeV
should be captured without sending the fireball. The excited
nucleus, created during such a process may deexcite by emission of
particles, in similar way as the heavy residuum of the intranuclear
cascade, but may also break up, what leads to the emission of two
excited sources of particles as it is in our picture of the reaction
mechanism.

The fireball emission and the break-up of the target, illustrated by
Figs. \ref{fig:firelow} and \ref{fig:break}, may appear at low
energy, e.g. 0.175 GeV, only exclusively.  It is important to
emphasize that, nevertheless, both are observed in the analyzed
data. This may be connected with the fact that at different impact
parameters the straight way of the proton through the target has
different length.  The peripheral collisions correspond to shorter
way through the nucleus, i.e., to the smaller stopping power,
whereas the central collisions, on the contrary, to the longest way
and the strongest stopping power. The estimation of Aichelin et al.
\cite{AIC84A} should be, thus,  treated as done for a specific case
- for central collisions.

The proton impinging with high kinetic energy, e.g., 1.2 GeV or
higher, can move through the nucleus knocking-out a fireball, but
relative momentum of the proton and the fireball may be so large,
that the proton flies independently of the fireball. Then the
momentum of the fireball is smaller than the momentum of the beam,
and it is not related to the beam momentum in an unambiguous manner.
This may be a reason why the velocity of the fireball, observed in
p+Ni collisions in the proton energy range from 1.2 to 2.5 GeV
\cite{BUD09A}, is smaller than that observed at 0.175 GeV. Moreover,
it does not change significantly, in contrast to strong variation of
the beam energy.

\begin{figure}
\begin{center}
\includegraphics[angle=0,width=0.5\textwidth]{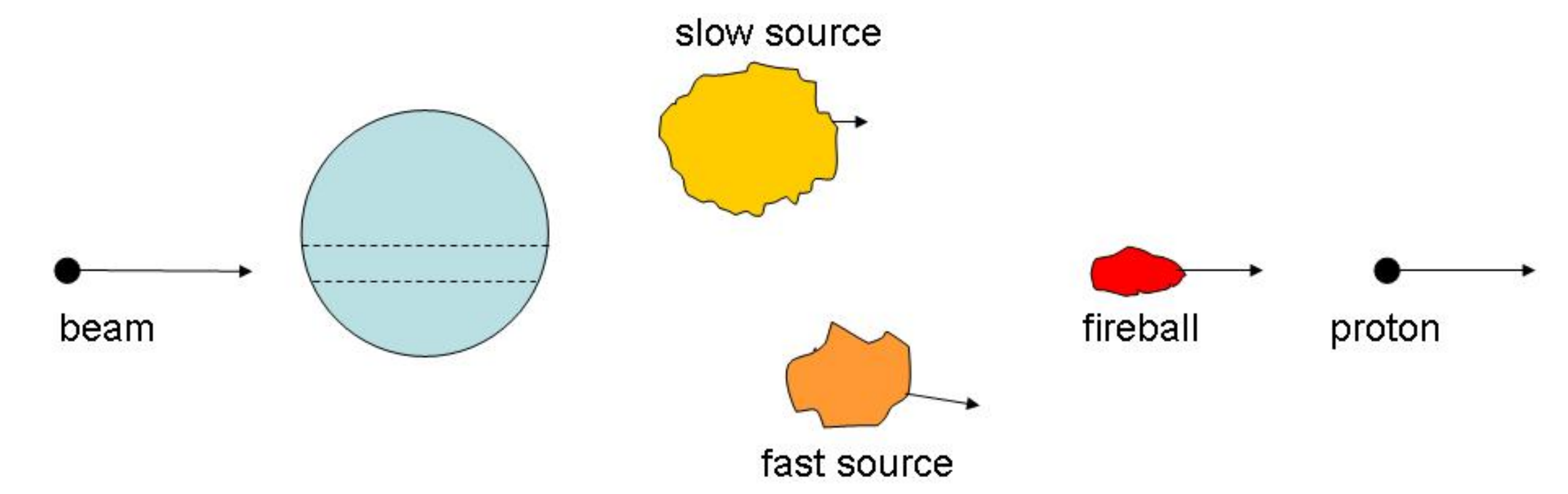}
\caption{\label{fig:firehigh} Fireball emission from Ni target at
high beam energy (over 1 GeV).}
\end{center}
\end{figure}

A schematic presentation of this reaction mechanism is shown in Fig.
\ref{fig:firehigh} for p+Ni collisions at energies above $\sim$ 1
GeV.

\begin{table}
\caption{\label{tab:lowhigh} Temperature and velocity parameters of
three sources of ejectiles for Ni target.  In the second column the
parameters obtained in the present study at a beam energy of 0.175
GeV are shown, whereas the third column presents parameters averaged
over beam energies 1.2, 1.9, and 2.5 GeV from \cite{BUD09A}. $T$
denotes apparent source temperature (in MeV), $\tau$ - the
temperature parameter corrected for the recoil, $A_S$ represents
mass number of the source, and $\beta$ its velocity in units of
speed of light. The symbol $A$ indicates the mass number of the
ejectile. Parameters with index $1$ correspond to slow source, with
index $2$ to fast source, and with index $3$ to the fireball. }
\begin{center}
 \begin{tabular}{c|c|c}
\hline \hline
   Parameter     &    Ni(0.175 GeV)                       &    Ni(1.2 - 2.5 GeV)  \\
\hline
 $T_1$           &    6.5(3)                 & 11.2(7) - 0.4(2) $\ast$ A \\
 $\tau_1$        &    6.5(3)                 & 11.2(7)   \\
 A$_{S_1}$       &     ?                     &  28(15)  \\
 $\beta_1$       &    [0.0036]               &  [0.005] \\
\hline
 $T_2$           & 11.2(1.0) - 0.25(17) $\ast$ A    & 22.5(6) - 0.8(1) $\ast$ A \\
 $\tau_2$        & 11.2(1.0)                 & 22.5(6) \\
 A$_{S_2}$       &   45(33)                  & 28(4) \\
 $\beta_2$       & 0.059(5) - 0.0034(6) $\ast$ A    & 0.044(6) - 0.0021(7) $\ast$ A \\
\hline
 $T_3$           & 28.1(2.3) - 6.6(1.4) $\ast$ A    & 52.7(1.1) - 9.6(4) $\ast$ A \\
 $\tau_3$        & 28.1(2.3)                 & 52.7(1.1) \\
 A$_{S_3}$       &  4.2(1.0)                 & 5.5(3) \\
 $\beta_3$       & 0.278(56) - 0.033(23) $\ast$ A   & 0.209(11) - 0.053(5) $\ast$ A   \\
 \hline
\end{tabular}
\end{center}
\end{table}

\section{Summary and conclusions}

The double differential cross sections $d^{2}\sigma/d\Omega dE$ for
the production of $^{1,2,3}$H, $^{3,4}$He and light IMFs (
$^{6,7}$Li, $^{7,9}$Be, $^{10,11}$B) in collisions of proton with a
Ni target hasve been measured at 0.175 GeV beam energy.  The aim of
the study was to investigate whether the nonequilibrium processes,
which were found to play an important role at higher energies (1.2,
1.9, and 2.5 GeV) \cite{BUD09A} are also present at such low energy
as 0.175 GeV.

The data were analyzed using a two-step microscopic model.  The
first step of the reaction was described as the intranuclear cascade
of
nucleon-nucleon 
collisions initiated by the proton from the beam.  It was allowed
that during the intranuclear cascade the coalescence of nucleons
into complex LCPs may proceed.  An emission of nucleons from the
cascade as well as an emission of LCPs created due to the
coalescence was the only nonequilibrium process taken explicitely
into consideration by this model.  The second step of the reaction
was assumed to be described as an evaporation of particles
(nucleons, LCPs and IMFs) from the equilibrated target residuum
after the intranuclear cascade.  It was found that main properties
of the spectra of LCPs are well reproduced by the model with the
exception of forward scattering angles in proton and deuteron
channels as well as all angles in the alpha particle channel.  The
IMF spectra were also not satisfactorily reproduced, especially for
high energy IMFs.

 It is
worthy to point out that a good quality of description of triton and
$^3$He spectra by coalescence of nucleons and evaporation was
achieved with  INCL4.3 and GEM2 computer programs, which used
default values of the parameters. Such a good agreement of
theoretical cross sections with the data measured at 0.175 GeV beam
energy is astonishing, because -- according to authors of the
program \cite{CUG97A,BOU04A} -- the present beam energy is on the
boarder of energy range where the concept of intranuclear cascade is
applicable.

A phenomenological analysis was performed assuming that additional
processes appear, which may be simulated by the emission from three
moving sources as it was successfully done at higher energies. It
was found that the cross sections of proton and deuteron emission
can be very well described when the emission from fireball, i.e.,
fast and hot source moving in the forward direction, was taken into
account.  The contribution of this process to the total production
cross sections is rather small - 20\% for protons and 14\% for
deuterons, however, it significantly improves description of the
spectra at forward angles.  Good reproduction of triton and $^3$He
spectra by two-step process, where the coalescence of nucleons was
very important, did not need any significant contribution of other
nonequilibrium processes.  The description of  alpha particle
spectra as well as IMF spectra was very improved by inclusion of
emission from two moving sources, which were interpreted as
prefragments of the target nucleus created due to break-up of this
nucleus caused by an impinging proton.

Due to a very good description of energy and angular dependencies of
differential cross sections $d^2\sigma/d\Omega dE$  it was possible
to extract total production cross sections for all investigated
ejectiles.  These cross sections are listed in Tables
\ref{tab:parameters} and \ref{tab:fireball} for IMFs and LCPs,
respectively.

The discussed above picture of the reaction mechanism agrees
generally with the picture of the nonequilibrium reactions
investigated at higher beam energies. There were, however, found
specific properties of the nonequilibrium reactions appearing at
0.175 GeV but not observed at high energies: (i) The fireball
exhausts the total available momentum, thus it cannot be accompanied
by break-up of the target remnant, 
(ii) The break-up of the target appears due to capture of the proton
without emission of the fireball.

\begin{acknowledgments}

The technical support of A.Heczko, W. Migda{\l}, and N. Paul in
preparation of experimental apparatus is greatly appreciated. This
work was supported by the European Commission through European
Community-Research Infrastructure Activity under FP6 "Structuring
the European Research Area" programme (Hadron Physics, contract
number RII3-CT-2004-506078 as well as the FP6 IP-EUROTRANS
FI6W-CT-2004-516520). One of us (MF) appreciates financial support
of Polish Ministry of Science and Higher Education (Grant No N N202
174735, contract number 1747/B/H03/2008/35). This work was also
partially supported by the Helmholtz Association through funds
provided to the virtual institute "Spin and strong QCD" (VH-VI-231).

\end{acknowledgments}

\end{document}